\documentclass[aps,amsmath,amssymb,nofootinbib,twocolumn]{revtex4}
\usepackage{latexsym}
\usepackage{graphics}
\usepackage{bm}
\usepackage{graphicx}
\usepackage{bm}
\usepackage{color}
\usepackage{amsmath}

\begin{document}

\newcommand{\be}{\begin{equation}}
\newcommand{\ee}{\end{equation}}
\newcommand{\ben}{\begin{eqnarray}}
\newcommand{\een}{\end{eqnarray}}
\newcommand{\nn}{\nonumber \\}
\newcommand{\ii}{\'{\i}}
\newcommand{\pp}{\prime}
\newcommand{\tr}{{\mathrm{Tr}}}
\newcommand{\nd}{{\noindent}}
\newcommand{\grad}{\hspace{-2mm}$\phantom{a}^{\circ}$}
\newcommand{\qunoinv}{\frac{1}{q-1}}
\newcommand{\quno}{q-1}


\title{Statistical manifestation of quantum correlations via disequilibrium}


\author{ F. Pennini$^{1,2}$, A. Plastino$^{3,4}$}

\affiliation{$^{1}$ Departamento de F\'{\i}sica, Universidad
Cat\'olica del Norte, Av.~Angamos~0610, Antofagasta,
Chile\\$^{2}$Departamento de F\'{\i}sica, Facultad de
Ciencias Exactas y Naturales, Universidad Nacional de La Pampa, CONICET, Av. Peru 151, 6300, Santa Rosa, La Pampa,
Argentina\\$^{3}$Instituto de F\'{\i}sica La Plata--CCT-CONICET,
Universidad Nacional de La Plata, C.C.~727, 1900, La Plata,
Argentina\\$^{4}$ SThAR - EPFL, Lausanne, Switzerland
 }

\date{\today}

\date{Received: date / Revised version: date}
\begin{abstract}
That of disequilibrium ($D$) is a statistical notion  introduced by L\'{o}pez-Ruiz, Mancini, and Calbet~(LMC)
 more than 20 years ago [Phys. Lett. A  {\bf 209} (1995) 321]. $D$ measures the amount of ``correlational structure'' of a system. We wish to use $D$ to analyze one of the simplest types of quantum correlations,
those present in simple quantum gaseous systems and due to  symmetry considerations.
 To this end we extend the LMC formalism to the grand canonical environment and show that
  $D$ displays distinctive  behaviors for simple gases, that allow for interesting insights into their structural properties.
\end{abstract} 


\date{\today}





\maketitle

\section{Introductory matters}
\label{Complex}

\subsection{ Historical notes}
Knowledge of the   unpredictability
and randomness of a system does not automatically translate in an adequate grasping of the extant
  correlation structures reflected by the current probability distribution~(PD). The desideratum
    is  to capture the
relations amongst  a system's  components in a similar manner as entropy describes disorder.
 One certainly knows that the antipodal  extreme cases of (a) perfect order and (b)
maximum randomness are not characterized by strong
correlations \cite{LMC}. Amidst  (a) and (b)  diverse correlation-degrees may be manifested  by  the features  of the probability distribution. The big question is how? Answering the query is not simple task.
Notoriously, Crutchfield has stated  in
1994 that \cite{Crutch1,Crutch2}:\vskip 2mm

``\textit{Physics does have the tools for detecting and measuring
complete order equilibria and fixed point or periodic behavior and
ideal randomness via temperature and thermodynamic entropy or, in
dynamical contexts, via the Shannon entropy rate and Kolmogorov
complexity. What is still needed, though, is a definition of
structure and a way to detect and to measure it \cite{Crutch1,Crutch2}}".\vskip 3mm \nd
 Famously, Seth Lloyd found as many as 40 ways of introducing a complexity definition,
none of which quite  satisfactory.\vskip 3mm \nd

LMC introduced an interesting  functional of the PD that does
grasp correlations in the way that  entropy
captures randomness. This may be regarded as a  great  breakthrough~\cite{LMC}. LMC's
statistical complexity  did individualize and quantify the bequeath of
Boltzmann's entropy (or information $H$) and that of structure. {\it The latter contribution came from the notion
of disequilibrium}. It measures (in probability space) the distance from i) the prevailing probability distribution  to ii) the uniform probability. {\it $D$ reveals the amount of structural details. The larger it is, the more structure exists}~\cite{LMC,cuatro}.  For  $N-$particles one has
\be
D=\sum_{i=1}^N\,\left(p_i-\frac{1}{N}\right)^2.
\ee
Here ${p_1, p_2,\ldots, p_N}$ are the individual normalized probabilities ($\sum_{i=1}^N\,p_i=1$)~\cite{LMC}. The two ingredients $H$ and $D$ are combined by LMC to yield the complexity $C$ in the fashion
$C_{LMC}=D H$~\cite{LMC,MPR,lmc1,lmc11,lmc2,lmc22,lmc3}). $C_{LMC}$
 vanishes, the two above extreme cases  (a) and (b).

\subsection{Our present task and its motivation}
In this paper we deal with  $D$ properties,  within a
grand canonical ensemble scenario, for simple gaseous system obeying quantum statistics. We  will  use $D$ as a structure-indicator so as to compare
the classical Maxwell-Boltzmann situation of no quantum correlations vis-a-vis the Bose-Einstein and Fermi-Dirac situations. \vskip 2mm

\nd  Why? Because in this way we have an opportunity of
 observing the workings of quantum symmetries in the simplest conceivable scenario.
We will indeed encounter interesting  quantum insights. \vskip 2mm

\nd The issue of separating quantum (qc) from classical correlations (cc) has revived since the discovery of quantum discord \cite{discord1,discord2,discord3}. Before, it was a simple matter to distinguish between cc and qc, because the former were associated to separable states and the latter to non-separable ones endowed with entanglement. The discovery that some separable states are also endowed with qc (discord) made the cc-qc distinction
a more formidable task, still the subject of much research. This gives our present endeavor some additional contemporary relevance.

\vskip 2mm

The structure of this paper is the following. In Section \ref{sec1} we recapitulate the relevant
background, i.e.,  the pertinent formalism in the canonical ensemble and  we introduce also our proposal referring to extending the disequilibrium  notion  to the grand canonical ensemble.
 The main results of the paper are in Section \ref{sec2}, in which we apply our ideas to  quantum gaseous systems, focusing attention  on the occupation number.
Finally, we present our conclusions en Section \ref{conclu}.

\section{Disequilibrium in the grand canonical ensemble}
\label{sec1}
\subsection{ L\'{o}pez-Ruiz work for the canonical ensemble}
\label{lmc}

We recapitulate first interesting   notions of
 L\'{o}pez-Ruiz, for the canonical ensemble \cite{cuatro,LRuiz2001}, dealing with a classical ideal gas in thermal
equilibrium. One has  $N$ identical particles, confined in a volume
$V$ at the temperature $T$. The ensuing
Boltzmann PD is~\cite{pathria1996}

\be \rho(x,p)=\frac{e^{-\beta
\mathcal{H}(x,p)}}{Q_N(V,T)}.\label{rho} \ee One has $\beta=1/k_BT$, $k_B$  Boltzmann's constant,
while $\mathcal{H}(x,p)$ is the Hamiltonian, and
$x,p$ the phase space variables. The
canonical partition reads

\be Q_N(V,T)=\int
\mathrm{d}\Omega\,e^{-\beta\,\mathcal{H}(x,p)},\label{parti}
\ee with $\mathrm{d}\Omega=\mathrm{d}^{3N} x\,
\mathrm{d}^{3N} p/N! h^{3 N }$. The
Helmholtz' free energy $A$ is~\cite{pathria1996}

\be A(N,V,T)=-k_B T\, \ln Q_N(V,T).\label{free} \ee R.
L\'{o}pez-Ruiz (LR)  demonstrates in Ref. \cite{LRuiz2001} that the
disequilibrium $D(N,V,T)$  displays the following nice appearance for continuous probability
distributions

\be D(N,V,T)=e^{2\beta\,[A(N,V,T)-A(N,V,T/2)]}.\label{deseq} \ee LR  changes now   $T$ by $ T/2$ in  Eq. (\ref{free}) and replaces this in Eq. (\ref{deseq}). Consequently, he finds

\be D(N,V,T)=\frac{Q_N(V,T/2)}{\left[{Q_N}(V,T)\right]^2}. \label{des0}\ee

\subsection{Our proposal for the grand canonical ensemble}

Our goal now is to extend the above LR-formulation to the grand canonical ensemble.
As it is well known, the natural quantity associated to this ensemble is the grand potential which is given by~\cite{pathria1996}

\be
\Psi(z,V,T)=-k_B T\,\ln \mathcal{Z}(z,V,T),\label{potentialG}
\ee
where $\mathcal{Z}(z,V,T)$ is the grand partition function of the system and the parameter $z$ is the fugacity  defined by $z=\exp(-\alpha)=\exp(\mu\beta)$~\cite{pathria1996}.
One can then extend the  ideas presented above for  the canonical ensemble, with the help of  the relationship between the free energy and the grand potential, that reads

\be
A(z,V,T)=N k_B T \ln z+ \Psi(z,V,T).
\ee

Introducing this into Eq. (\ref{deseq}), one re-express the disequilibrium in the grand canonical ensemble which is now
\be
D(z,V,T)=e^{2\beta\,[\Psi(z,V,T)-\Psi(z^2,V,T/2)]}.\label{1deseqGC}
\ee
 Note we are adding a dependence on the fugacity $z$ in $D$. We observe that, when $T$ changes to $T/2$, then $z$ is replaced by $z^2$.
Thus, using  Eq. (\ref{potentialG}), we immediately are led to an original (we believe)  disequilibrium
expression in terms of the grand partition function
\be
D(z,V,T)=\frac{\mathcal{Z}(z^2,V,T/2)}{{\mathcal{Z}}^2(z,V,T)}, \label{2desGC}
\ee
which,  depends on the variables $V$, $T$, and fugacity $z$.


\section{Statistical features of the quantum gaseous system' occupation number}
\label{sec2}
\subsection{Disequilibrium}

Following Ref.~\cite{pathria1996}, Chapter 6, we focus attention on a gaseous system of $N$ non-interacting undistinguishable particles contained in a volume~$V$ with energies $\epsilon_k$ grouped into cells  as described in this classical book.  In the grand canonical ensemble, the equation of state for the aforementioned system, is given
by~\cite{pathria1996}

\be
\frac{P  V}{k_B T}=\ln \mathcal{Z}(z,V,T)=\frac{1}{a}\sum_{\epsilon}\,\ln(1+a z e^{-\beta \epsilon}),\label{stategc}
\ee
where $a=+1$ in the Fermi-Dirac (FD) case, $a=-1$ in the Bose-Einstein (BE) one, and $a=0$ for the Maxwell-Boltzmann (MB)
instance. The energy $\epsilon$  runs over  every eigenstate. In particular, for the
classical case, the grand partitions functions becomes~\cite{pathria1996}

\be
\mathcal{Z}(z,V,T)=z \sum_{\epsilon} e^{-\beta \epsilon}.\label{za0}
\ee

Replacing Eqs. (\ref{stategc}) and (\ref{za0}) into Eq. (\ref{2desGC}) we analytically get the disequilibrium whose form is

\be
D(z,V,T)=\prod_{\epsilon}\,D_{a}(z,\epsilon, V,T),
\ee
where, for each energy level, we have

\be
D_{a}(z,\epsilon, V, T)= \left\{ \begin{array}{ll}
\frac{(1+a z^2 e^{-2\beta \epsilon})^{1/a}}{(1+a z e^{-\beta \epsilon})^{2/a}}&\,\,\, \textrm{for $a=\pm 1$},\label{dis1}\\\\
z e^{-\beta \epsilon}&\,\,\, \textrm{ for $a=0$},
\end{array} \right.
\ee
a disequilibrium  expression for  the level of energy  $\epsilon$ for the three cases under consideration. From hereafter, in order to simplify the notation, we will drop the variables $z$, $V$, and $T$. Therefore,  only the dependency on $\epsilon$ will be preserved.

Moreover, since the mean occupation number $\langle n_{\epsilon}\rangle$ of the level $\epsilon$ is given by~\cite{pathria1996}

\be
\langle n_{\epsilon}\rangle=\frac{1}{z^{-1}e^{\beta\epsilon}+a},\label{mean}
\ee
then we immediately have that

\be
z^{-1}e^{\beta\epsilon}=\frac{1}{\langle n_{\epsilon}\rangle}-a.\label{factor}
\ee
Therefore, replacing Eq. (\ref{factor}) into the couple of Eqs. (\ref{dis1})  we obtain $D_{a}(\epsilon)$ as a function of the occupation number for the three cases. This reads

\be
D_{a}(\epsilon)= \left\{ \begin{array}{ll}
\left[(1-a \langle n_{\epsilon}\rangle)^2+a \langle n_{\epsilon}\rangle^2\right]^{1/a}&\,\,\, \textrm{for $a=\pm 1$},\label{dis3}\\\\
\langle n_{\epsilon}\rangle &\,\,\, \textrm{ for $a=0$}.
\end{array} \right.
\ee

\begin{figure}[h]
\begin{center}
\includegraphics[scale=0.6,angle=0]{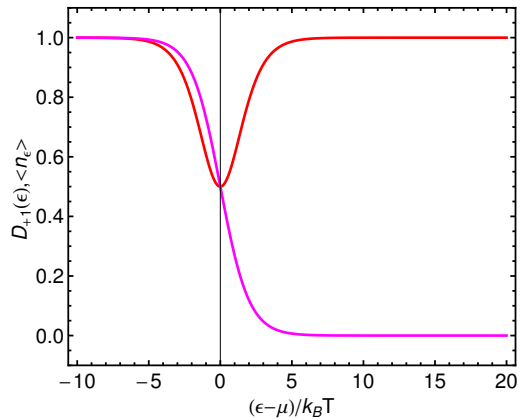}
\vspace{-0.2cm} \caption{ Disequilibrium  $D_{+1}(\epsilon)$ (red curve) and $\langle n_{\epsilon}\rangle$ (magenta curve) versus $(\epsilon-\mu)/k_B T$ for fermions.  }\label{figeps1}
\end{center}
\end{figure}

\begin{figure}[h]
\begin{center}
 \includegraphics[scale=0.6,angle=0]{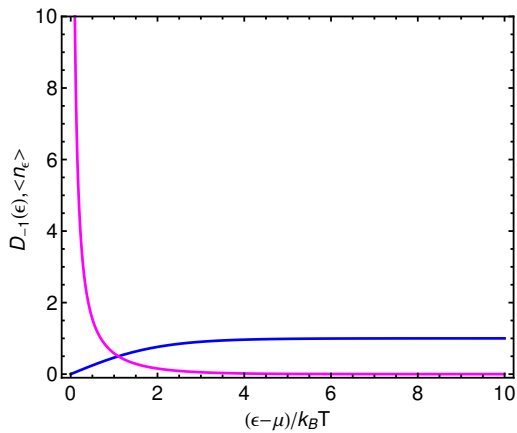}
\vspace{-0.2cm} \caption{ Disequilibrium  $D_{-1}(\epsilon)$ (blue curve) and $\langle n_{\epsilon}\rangle$ (magenta curve) versus $(\epsilon-\mu)/k_B T$ for bosons.  }\label{figeps2}
\end{center}
\end{figure}

\begin{figure}[h]
\begin{center}
\includegraphics[scale=0.6,angle=0]{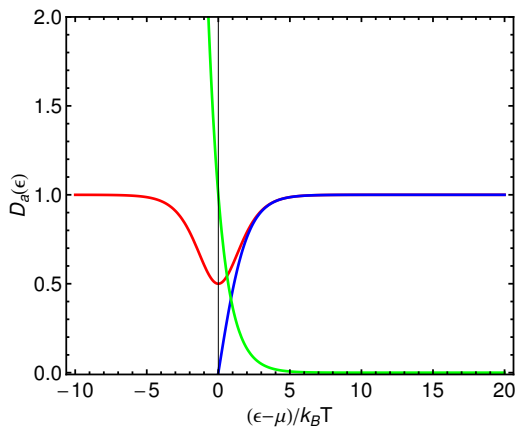}
\vspace{-0.2cm} \caption{Disequilibrium  $D_{a}(\epsilon)$   versus  $ (\epsilon-\mu)/k_B T $.  We label red curve for fermions, blue line for bosons and green curve for MB (no quantum correlations).  }\label{figu31}
\end{center}
\end{figure}

\begin{figure}[h]
\begin{center}
\includegraphics[scale=0.6,angle=0]{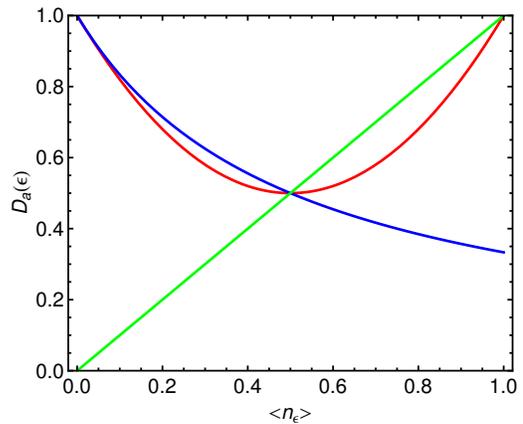}
\vspace{-0.2cm} \caption{ Disequilibrium  $D_{a}(\epsilon)$   versus  $\langle n_{\epsilon}\rangle$. We label red curve for fermions, blue line for bosons and green curve for MB (no quantum correlations).  }\label{figu32}
\end{center}
\end{figure}

The $D_{a}(\epsilon)-$behavior  ruled by Eqs. (\ref{dis1}) and (\ref{dis3}) is displayed in Figs.~\ref{figeps1} to \ref{figu32} by the FD (red),  BE (blue), and MB (green) cases. The differences with the  classical result are a magnificent illustration of quantum correlations. The minimum of $D_{a}(\epsilon)$ occurs when $\langle n_{\epsilon} \rangle=\,\,  1/(1+a)$, i.e., $\langle n_{\epsilon} \rangle=1/2$ for fermions and $\langle n_{\epsilon} \rangle=\infty$ for bosons, as we illustrated in Figs.~\ref{figeps1} and \ref{figeps2}.  Minimum $D_{a}(\epsilon)$ entails minimal structure, that in the BE instance is associated to the condensate. Thus, the condensate is endowed with minimum structure, i.e.,
$D_{-1}(\epsilon)$ clearly identifies the condensate {\it as having no
distinctive {\bf structural} } features, which constitutes a new $D_{a}(\epsilon)-$result, as far as we know. This notion is reinforced in Figs.~\ref{figu31} and \ref{figu32}, by plotting $D_{a}(\epsilon)$ versus $\langle n_{\epsilon} \rangle$.
If we set $q=(\epsilon-\mu)/k_B T$, we appreciate the fact that $D_{a}(\epsilon)$ varies only between $q=0$ and approximately $q=10$, remaining constant equal to 1 for any larger $q-$value grater than 10. Remember that the simple gaseous systems' descriptions converge to the classical one as $q$ grows \cite{pathria1996}.
\vskip 3mm

\nd In the FD instance, instead, the minimum of $D_{+1}(\epsilon)$ obtains for the situation farthest removed from the trivial instances of zero or maximal occupation.   For fermions, complete or zero occupations display maximal structure. At fist sight, this
 behavior near $\langle n_{\epsilon}\rangle=0$ may seem surprising. The quantum disequilibrium is large while the classical one vanishes.  There is no structure, classically. However, it is well known that the quantum vacuum is a very complex, complicated object, as quantum electrodynamics clearly shows (the quantum-vacuum literature is immense. A suitable introductory treatise is that of Mattuck in Ref.~\cite{mattuck}).  This is foreshadowed by the quantum disequilibrium at the level of quantum gases! Instead, we note that, for $a=0$, (the classical case), $D_{0}(\epsilon)$ coincides with the mean occupation number.

\vskip 3mm \nd On the other hand, let us reiterate that for $\langle n_{\epsilon}\rangle \rightarrow \infty$, the boson disequilibrium vanishes, on account of dealing with indistinguishable particles. The condensate exhibits no structural details.  Instead, the MB $D_{0}(\epsilon)$ grows with $\langle n_{\epsilon}\rangle$ because one deals with distinguishable particles, and much more information is needed to label a million particles than 10 of them. This fact emphasizes the fact that  $D_{a}(\epsilon)$  tells us about {\it information} on structural details, either physical or  labeling-ones.

\begin{figure}[h]
\begin{center}
\includegraphics[scale=0.6,angle=0]{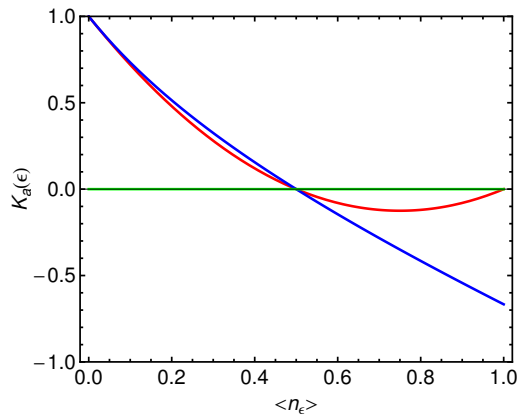}
\vspace{-0.2cm} \caption{$K_{a}(\epsilon)=D_{a}(\epsilon) - \langle n_{\epsilon}\rangle$ as a function of  $\langle n_{\epsilon}\rangle$ (red curve for fermions, blue curve for bosons and green curve for MB), the quantumness indicator. }\label{qnesss}
\end{center}
\end{figure}

\vskip 3mm \nd Let us define $K_{a}(\epsilon)=D_{a}(\epsilon) - \langle n_{\epsilon}\rangle$, which is a $D_{a}(\epsilon)-$related
``quantumness index", given that it vanishes in the classical case for all mean occupation number. We plot it in Fig. \ref{qnesss}. This graph is very instructive.
Note that $\langle n_{\epsilon}\rangle=1/2$ is a  critical value.
For it, the curves attain classical values and, for fermions,  $D_{a}(\epsilon)$ is minimum, reflecting on minimal fermion structure. Not surprisingly, in view of previous considerations, $K_{a}(\epsilon)$ is maximal at the quantum vacuum. From  $\langle n_{\epsilon}\rangle=0$, $K_{a}(\epsilon)$~steadily diminishes till we reach the critical point mentioned  above. For bosons, it then steadily increases again, in absolute value, towards the condensate.  For fermions it grows again, in absolute value, reaches a maximum at   $\langle n_{\epsilon}\rangle =3/4$,   and then tends
to zero again at $\langle n_{\epsilon}\rangle =1$.

\subsection{Probability distributions as a function of $D_{a}(\epsilon)$}

It is well-known the the probability to encounter  exactly $n$ particles in a state of
energy $\epsilon$ is $p_{\epsilon}(n)$~\cite{pathria1996}, which for the Fermi-Dirac instance reads

\be
p_{\epsilon}(n)|_{FD}= \left\{ \begin{array}{ll}
1- \langle n_{\epsilon}\rangle&\,\,\, \textrm{for $n=0$},\label{proba}\\\\
\langle n_{\epsilon}\rangle &\,\,\, \textrm{ for $n=1$}.
\end{array} \right.
\ee
Thus, considering Eqs. (\ref{dis3}) and (\ref{proba}), the disequilibrium becomes

\be
D_{+1}(\epsilon)=\sum_{n=0}^{1}\,p_{\epsilon}^2(n)=p_{\epsilon}^2(0)+p_{\epsilon}^2(1).
\ee
Since $p_{\epsilon}(0)+p_{\epsilon}(1)=1$, replacing this into above equation, we also have

\be
D_{+1}(\epsilon)=(1-p_{\epsilon}(1))^2+p_{\epsilon}^2(1),\label{dpfermi}
\ee
the disequilibrium as a function of the probability of the occupation of the level of energy $\epsilon$ with one fermion.

Solving Eq. (\ref{dpfermi}) we get

\be
p_{\epsilon}(1)|_{FD}=\frac12 (1\pm\sqrt{2 D_{+1}(\epsilon)-1}),\label{2dpfermi}
\ee
which leads to bi-valuation in expressing probabilities as a function of $D_{+1}$, a novel situation uncovered here.
\vskip 3mm

\nd In the Bose-Einstein case, the probability is the geometrical distribution~\cite{pathria1996}
\be
p_{\epsilon}(n)|_{BE}=\frac{\langle n_{\epsilon}\rangle^n}{(1+\langle n_{\epsilon}\rangle)^{n+1}},
\ee
and, accordingly, the disequilibrium is now of the form
\be
D_{-1}(\epsilon)=\frac{1-p_{\epsilon}(n)}{1+p_{\epsilon}(n)}.\label{dpbose}
\ee
From the above equation then we get the probability distribution as a function of the disequilibrium that it reads as follows
\be
p_{\epsilon}(n)_{BE}=\frac{1-D_{-1}(\epsilon)}{1+D_{-1}(\epsilon)}.\label{2dpbose}
\ee

\vskip 3mm

\nd For the MB-instance, $p_{\epsilon}$ is a Poisson distribution  given by~\cite{pathria1996}

\be
p_{\epsilon}(n)|_{MB}=\frac{(\langle n_{\epsilon}\rangle)^{n}}{n!} e^{-\langle n_{\epsilon}\rangle},
\ee
that, for Eq. (\ref{dis3}) becomes

\be
p_{\epsilon}(n)|_{MB}=\frac{D_{0}(\epsilon)^{n}}{n!}\, e^{-D_{0}(\epsilon)}.\label{dpMB}
\ee

\begin{figure}[h]
\begin{center}
\includegraphics[scale=0.6,angle=0]{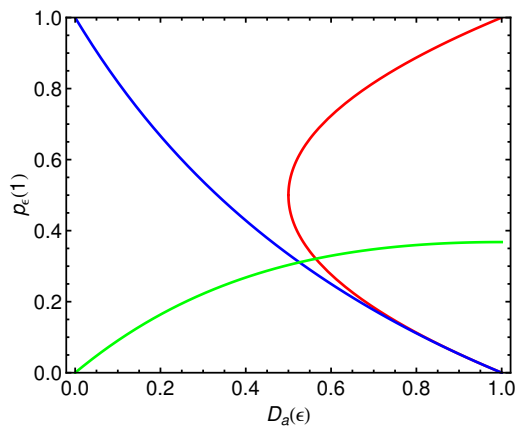}
\vspace{-0.2cm} \caption{Probability $p_{\epsilon}(1)$  versus  $D_{a}(\epsilon)$ for three cases mentioned.  We label red curve for fermions, blue curve for bosons and green curve for MB (no quantum correlations).}\label{probas}
\end{center}
\end{figure}

We represent Eqs. (\ref{2dpfermi}), (\ref{2dpbose}) and (\ref{dpMB}) in Fig. \ref{probas}, where we plot the probability as a function of the disequilibrium $D_{a}(\epsilon)$ for the three cases here discussed, with $a=0, \pm 1$. All of them are distinctly different. The classical one is a Poisson distribution. The boson-one decreases steadily as $D_{-1}(\epsilon)$ augments.
The FD distribution is bi-valuated save at the $D_{+1}(\epsilon)=1/2$. Note that for $D_{+1}(\epsilon)=1$  the probability can be either zero or one, a kind of ``cat"-effect.

\subsection{Disequilibrium in term of fluctuations}
\begin{figure}[h]
\begin{center}
\includegraphics[scale=0.6,angle=0]{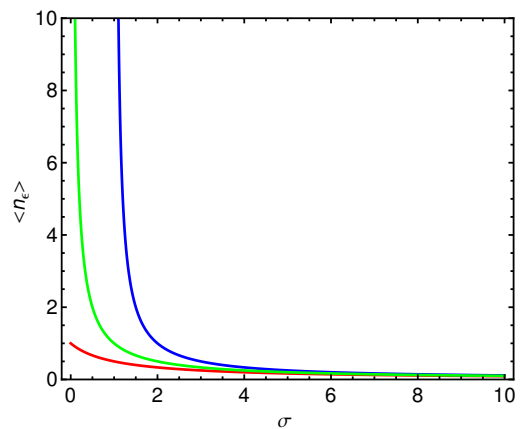}
\vspace{-0.2cm} \caption{The mean occupation number $\langle n_{\epsilon} \rangle$  versus~$\sigma$.}\label{flunmedio1}
\end{center}
\end{figure}

\begin{figure}[h]
\begin{center}
\includegraphics[scale=0.6,angle=0]{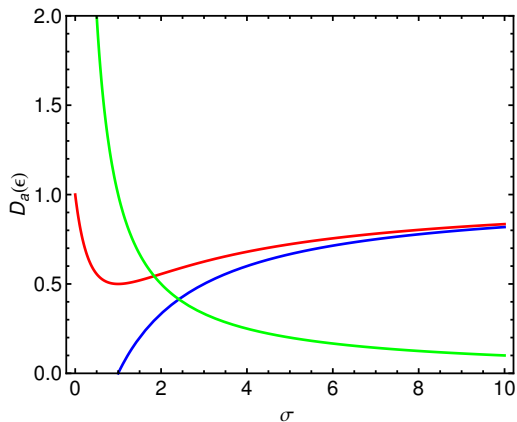}
\vspace{-0.2cm} \caption{Disequilibrium $D_{a}(\epsilon)$  versus~$\sigma$. We label red curve for fermions, blue curve for bosons and green curve for MB (no quantum correlations).}\label{flunmedio2}
\end{center}
\end{figure}

The relative mean-square fluctuation is~\cite{pathria1996}

\be
\sigma=\frac{\langle n_{\epsilon}^2\rangle-\langle n_{\epsilon}\rangle^2}{\langle n_{\epsilon}\rangle^2}=\frac{1}{\langle n_{\epsilon}\rangle}-a.
\ee
In the classical case ($a=0$) the relative fluctuation is ``normal", in the sense that it is proportional to the inverse occupation number and exhibits statistical behavior of uncorrelated events. In the Fermi-Dirac case $\sigma$ becomes subnormal and fermions exhibit a negative statistical correlation. Since $0\leq\langle n_{\epsilon}\rangle\leq 1$, then $\sigma\geq 0$. On the other hand, in the Bose-Einstein case, the fluctuation is super-normal~\cite{pathria1996} ($\sigma \geq 1$), thus bosons exhibit positive statistical correlation~\cite{pathria1996}.  We observe all this in Figs. \ref{flunmedio1} and \ref{flunmedio2}.

Therefore, in terms of the fluctuations we get

\be
D_{a}(\sigma)= \left\{ \begin{array}{ll}
\left(\frac{a+\sigma^2}{(a+\sigma)^2}\right)^{1/a}& \textrm{for $a=\pm 1$},\label{disigma}\\\\
1/\sigma & \textrm{ for $a=0$}.
\end{array} \right.
\ee
We see that, in quantum terms, $D_{a}(\sigma)$ strongly depends also upon the symmetry parameter $a$.
We plot the couple of Eqs.~(\ref{disigma}) in Fig.~\ref{flunmedio1}. As $\sigma$ grows, fermion's and boson's $D_{a}(\sigma)$s tend to coincide.
 Once again, for fermions the normal situation is that of minimum $D_{a}(\sigma)$.


\section{Conclusions}
\label{conclu}

\nd We have shown in this note that quantum effects are clearly reflected by the disequilibrium's $D_{a}(\epsilon)$ behavior.

\begin{itemize}
 \item    For instance, minimum $D_{-1}(\epsilon)$ entails minimal structural correlations, that in the BE instance are associated to the condensate. Thus, the condensate is endowed with minimum structure, i.e.,
$D_{-1}(\epsilon)$ clearly identifies the condensate as having no distinctive structural features. This  is reinforced in Fig.~\ref{figu32}, by plotting $D_{-1}(\epsilon)$ versus~$\langle n_{\epsilon} \rangle$.
\vskip 3mm

 \item  On the other hand, in the FD instance the minimum of $D_{+1}(\epsilon)$ obtains for the situation farthest removed from the trivial instances of zero or maximal occupation.   For fermions, complete or zero occupations display maximal structure.

 \item    The behavior near $\langle n_{\epsilon}\rangle=0$ is remarkable. The quantum disequilibrium is large while the classical one vanishes.  There is no structure, classically. However,  the quantum vacuum is a very , complicated object, as quantum electrodynamics clearly shows.  This is foreshadowed by the quantum disequilibrium at the level of simple gaseous systems.

\ \item    For $\langle n_{\epsilon}\rangle \rightarrow \infty$, the boson disequilibrium vanishes, on account of dealing with indistinguishable particles. The condensate exhibits no structural details.
on the contrary, the MB $D_{0}(\epsilon)$ grows with $\langle n_{\epsilon}\rangle$ because one deals with distinguishable particles, and much more information is needed to label a million particles than 10 of them.

 \item    We then gather  that  $D_a(\epsilon)$  tells us about {\it information} on structural details, either physical or  labeling-ones.

\end{itemize}

\end{document}